\documentclass[twocolumn,%preprint
,showpacs, showkeys, preprintnumbers,amsmath,amssymb, prb, aps]{revtex4}
\usepackage{dcolumn}
\usepackage{bm}
\usepackage[dvips]{graphicx}
\begin{document}
\title{Controllable Josephson current through a pseudo-spin-valve structure}
\author{C. Bell, G. Burnell, C. W. Leung, E. J. Tarte, D.-J. Kang, M. G. Blamire}
\affiliation{Materials Science Department, IRC in Superconductivity and IRC in Nanotechnology, \\
University of Cambridge, United Kingdom}
\date{\today}
\begin{abstract}
A thin Co/Cu/Permalloy (Ni$_{80}$Fe$_{20}$) pseudo-spin-valve structure is sandwiched between
superconducting Nb contacts. When the current is passed
perpendicular to the plane of the film a 
Josephson critical current ($I_C$) is observed at 4.2 K, in addition to a
magnetoresistance (MR) of $\sim$ 0.5 \% at high bias. The hysteresis loop of the
spin-valve structure can be cycled to modulate the zero field $I_C$ of
the junction in line with the MR measurements. These modulations of
resistance and $I_C$ occur both smoothly and sharply with the applied
field. For each type of behaviour there is a strong correlation
between shape of the MR loops and the $I_C$ modulation.
\end{abstract}
\pacs{74.50+r, 75.47.De, 85.25.Cp, 85.75.-d}
\keywords{Josephson junction, pseudo-spin-valve}
\maketitle
Josephson junctions with ferromagnetic barriers (SFS) have been 
investigated recently using weak ferromagnets.  Due to the oscillating superconducting order parameter in the
F layer, the groundstate phase difference of the S electrodes was
found to change from 0 to $\pi$ (a so-called $\pi$-junction) as a
function of temperature and F thickness \cite{ryazanov1, kontos}. SFS junctions have also been fabricated
with the strong ferromagnets Ni \cite{blum} and Co \cite{surgers}. In
the above cases the F layer has a profound effect on the
properties of the junction. Using an F layer to controllably alter the
properties of a Josephson junction is important for a 
fundamental understanding of the physics, but also has potential
applications as a memory element.

In this Letter we extend the work on single F layer junctions and
examine a S/pseudo-spin-valve (PSV)/S Josephson junction. The PSV
consisted of two F layers with different coercive fields, separated by
a Cu spacer. This artifical structure allows active control of the
magnetic state of the barrier. In contrast to previous
relatively large SFS junctions where the F layer was demagnetised,
here we actively pursue inhomogeneous magnetic structure. The
magnetoresistance (MR) of the PSV gives direct access to information
about relative orientation of the Co and Py layers, and 
the magnetic state of the barrier. The junction dimensions were such that the
field $H$ required to switch the F layers at 4.2 K is comparable to the
$I_C (H)$ modulation period. The PSV
structure with no antiferromagnetic pinning layer avoids the
strong reduction of critical current density $J_C$ found in for example
$\gamma$-FeMn Josephson junctions \cite{bellprb}. We will
show that $I_C$ can be controlled by putting the PSV into the
parallel and anti-parallel remenant states. 

Nb/Py/Cu/Co/Nb films were deposited on (100) oxidised silicon substrates by
d.c. magnetron sputtering at 0.5 Pa, in an in-plane magnetic field
$\mu_0 H \sim 40$ mT. The sputtering system was cooled with liquid
nitrogen and had a base pressure better than $3 \times
10^{-9}$ mbar. The Nb thicknesses were 180 nm and Cu
spacer thickness was 8 nm, (to avoid significant `orange
peel' magneto-static coupling between F layers). The Co and Py layers
which showed a Josephson current
were of the order of 1 nm and 1.6 nm respectively. The
films were patterned to micron scale wires with broad beam Ar ion
milling (1 mAcm$^{-2}$, 500 V), and then processed with a Ga focused ion
beam to achieve vertical transport with a device area in the
range $0.2 - 1$ $\mu$m$^{2}$. The fabrication process is
described in detail elsewhere \cite{bell}. 

\begin{figure}[h]
\includegraphics[width=8.5cm]{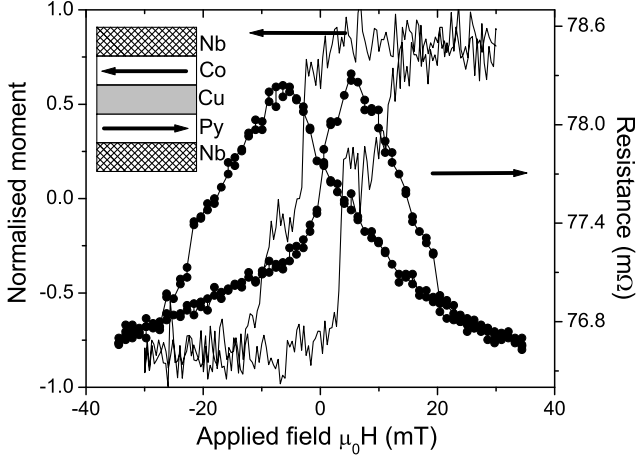}
\caption{\label{20165compare}Comparison of $M(H)$
loop at 30 K for a film with relatively thick F layers (2 nm Co and 3.2 nm Py),
with a $R(H)$ measurement (symbols) at 4.2 K. MR $\sim$ 2\% in this case. Inset:
schematic of the device (not to scale).}
\end{figure}

\begin{figure}[h]
\includegraphics[width=8.5cm]{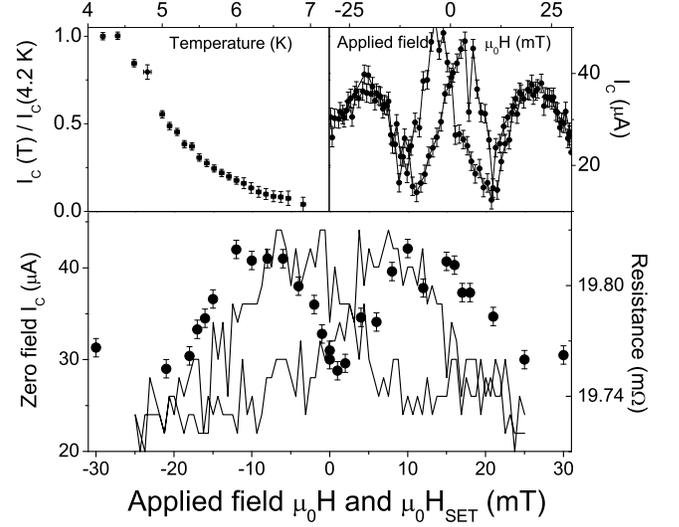}
\caption{\label{smoothspin}Smooth variation of the zero
field $I_C$ (symbols) compared to $R(H)$ (solid line). Junction size was
950 $\times$ 570 nm. Left inset: $I_C (T)$ normalised to $I_C$(4.2 K). Right inset: Hysteretic $I_C(H)$ for a 320 $\times$ 780 nm device.}
\end{figure}

\begin{figure}[h]
\includegraphics[width=8.5cm]{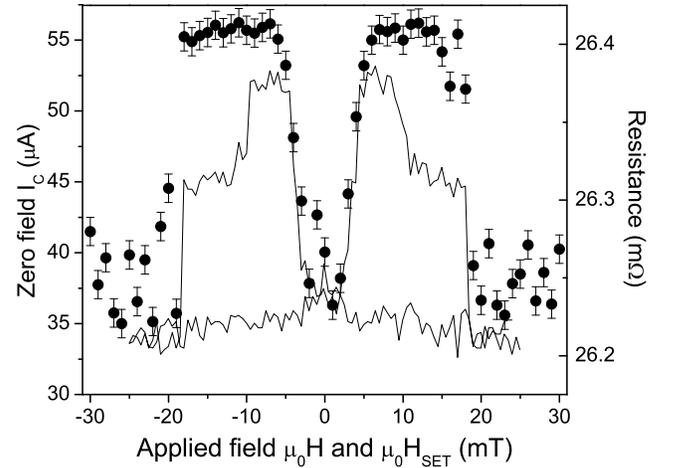}
\caption{\label{suddenspinagain}Sudden jump of zero field $I_C$ (symbols)
compared to $R(H)$ (solid line). Junction is 600 $\times$ 730 nm.}
\end{figure}

Transport measurements were made in a liquid He dip probe.  For the $I_C (H)$
and $R(H)$ measurements a quasi-d.c. ($\sim 15$ Hz) $IV$ was directly measured and $I_C$ extracted
using a voltage criterion. The $R(H)$ measurements 
used a bias current of $2-3$ mA, which is much
greater than $I_C$, but less than the critical current of the Nb
electrodes. The magnetic field was applied in-plane in the direction of the
applied field during deposition. As seen in the relatively thick device in Fig. \ref{20165compare}, the $R(H)$
and $M(H)$ measurements are consistent with one another as
expected. For the thinner F layers, Figs. \ref{smoothspin} and \ref{suddenspinagain} show the two different
behaviours of $R(H)$. The sudden jumps in Fig. \ref{suddenspinagain} compared to
the smooth variation of $R$ in Fig. \ref{smoothspin} can be explained by
sudden switches of a small number of domains in the barrier. The crossover from
smooth to sharp $R(H)$ switching takes place as the junction area $A$
is reduced below $\sim$ 0.45 $\mu$m$^2$, but is not strongly dependent
on the device aspect ratio, (the
aspect ratio is the junction dimension perpendicular to the
deposition field / dimension parallel to it). In both cases, despite the
different $R(H)$ behaviour, MR = ($R_{\mathrm{max}}$-$R_{\mathrm{min}}$) /
$R_{\mathrm{max}}$ of the order of 0.5 \% was obtained for the devices
in this work. The MR is consistent with the F layer and spacer thicknesses \cite{slater}.  The $I_C (H)$ is not a
`sinc' function as seen in uniform junctions,
and is hysteretic in the same sense as the MR measurements (inset of
Fig. \ref{smoothspin}). In this case the junction dimension in direction
perpendicular to the field is 320 nm. The first minima of the $I_C
(H)$ pattern for this junction dimension with a non-magnetic barrier
would be expected at $\mu_0 H$ $\sim$ 60 mT:\cite{bellprb} hence
there is extra complication added by the magnetic
barrier.

\begin{figure}[h]
\includegraphics[width=8.5cm]{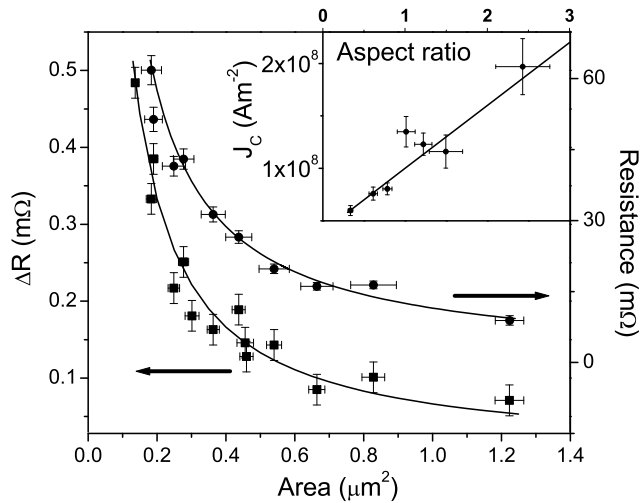}
\caption{\label{scalings}Scaling of $R$ and $\Delta R$
with $A$, lines are best fit to 1/$A$ power law. Inset: scaling of $J_C$ with aspect ratio.}
\end{figure}

To avoid this added complexity 
`zero field' $I_C$ was also measured. Hence the device was
prepared in distinct remenant states. To achieve this a
saturating field of $\pm$ 30 mT was
first applied. A field $\mp H_{\mathrm{SET}}$, was then applied in the opposite direction. The
field was then reduced to zero. The differential
resistance as a function of bias current of the junction was then made with a lock-in
amplifier, and the $I_C$ found. In
this way the parallel and anti-parallel remenant state $I_C$s were
measured. Hence `applied field' in Figs. \ref{smoothspin} and
\ref{suddenspinagain} for the `zero field $I_C$' is $H_{\mathrm{SET}}$. `$I_C$' is the average of the
absolute values of the positive and negative $I_C$s. The symbols in Figs. \ref{smoothspin} and
\ref{suddenspinagain} show zero field $I_C$.  The correlation between 
$I_C$ and $R(H)$ is striking. The first sudden switch of $R(H)$ in
Fig. \ref{suddenspinagain} but not in $I_C$ is attributed to a metastable state not 
present at zero field. For several devices $\Delta I_C$ = ($I_C^{\mathrm{max}}$-$I_C^{\mathrm{min}}$) /
$I_C^{\mathrm{max}}$ $\sim$ 30 \%, extremal values of 17 \% and 45 \% were also
obtained, but there was no scaling with dimension, area or aspect
ratio. The $I_C (T)$ monotonically decreased with $T$,
with parabolic behaviour as $I_C \rightarrow 0$, (inset Fig. \ref{smoothspin}). When normalised to
the $I_C$ at 4.2 K both remenant states showed very similar behaviour:
we attribute this to the $T$ dependence of the superconducting gap.
The change in $R$, $\Delta R$ and $R$ were inversely proportional to $A$
(Fig. \ref{scalings}): hence interface scattering is the dominant mechanism of MR \cite{albert}. Current
induced switching is not present since the
current density is $\sim$ 1 $\times$ $10^{10}$ Am$^{-2}$ which is a few orders of magnitude lower than required
\cite{albert}. $A\Delta R$ was in the range $5.4 - 8.7 \times 10^{-17}$ $\Omega$m$^2$,
and $AR = 1.1 \pm 0.2 \times 10^{-14}$ $\Omega$m$^2$. Using the two-current model from Yang
et al \cite{yang}, with their `best fit' parameters for Nb, Cu, Co and
Py, for our thicknesses $AR = 9 \pm 1 \times
10^{-15}$ $\Omega$m$^2$ in good agreement with our results. $A\Delta R$ was too small to
predict within the errors of the parameters of Yang et al. 
The junction $R_N$ is the same as the
value of the $R(H)$ measurements, and  will be referred to as simply $R$.
The $I_C R$ was in the range 0.8 to 2 $\mu$V. The $I_C R$ for a
given thickness of barrier should be constant. Hence despite the
apparent large difference in the shape of the $R(H)$, the $I_C R$ is
the same to within a
factor of approximately two, which is reasonable allowing for a few \AA
variation of barrier thickness over the chip. Given that Co and Py are
strong ferromagnets, we expect a small $J_C$,
despite the thin barrier thickness. Compared to the Co
junctions of S{\"u}rgers et al \cite{surgers}, where the barrier thickness was
5 nm, $J_C$ at 2.1 K was $\sim$ 1 $\times$ $10^7$
Am$^{-2}$. Assuming $J_C \propto \exp(-d/\xi_F )$ with $\xi_F$
$\approx$ 1 nm, a thickness 2 nm
gives $J_C$ $\sim$ 2 $\times$ $10^7$ Am$^{-2}$. Although the
comparison is crude, the experimental values in the range $0.5 - 1.2$ $\times$
$10^8$ Am$^{-2}$ are somewhat larger than this. $J_C$ is constant to within a
factor of two over the range of areas, but showed an interesting scaling with
device aspect ratio (inset of Fig. \ref{scalings}), which is not
understood at present. The value of $I_C$ and $R(H)$ behaviour with thermal cycling is not
perfectly reproducible, but show the same qualitative behaviour. This
is possibly due to different magnetic configurations `frozen in' when cooled. 

With the device geometry stray flux from the electrodes is present
\cite{bell}. However given the strong correlation
between $I_C$ and $R$ for the two very different
types of $R(H)$ behaviour, and that $R$ is determined solely by
magnetic structure of the barrier, we
conclude that the effect of stray field from the electrodes is not
important in directly determining $I_C$. A more quantitative
argument can be used \cite{clinton2}:
modelling the fringe field as a line source, the 
flux density $B =\mu_0 M_S d_F /2\pi r$, where $r$ is the distance from the
end of the track, and $d_F$ the F thickness. The saturation
magnetisation $M_S = 1.42 \times 10^6$ Am$^{-1}$ for Co. 
Considering the barrier as simply a 2 nm thick Co,
for $r$ = 500 nm, $B$ $\sim$ 1 mT. 
The magnetic induction from the barrier itself must also
be considered.  Using a simplistic model, in the worst case if we
assume all of the moment present in the barrier passes into the junction, 
perpendicular to the dimension $x$, then the flux
in the junction is $\Phi =xd_F \mu_0 M_S$. For $x
= 0.5$ $\mu$m, $\Phi$ $\sim 0.8 \Phi_0$, ($\Phi_0 = h/2e$ is the flux
quantum). This can
explain the suppression of $I_C$, but not the lack of scaling of $\Delta I_C$ with 
the dimension $x$, and also neglects the `return flux' in the
opposite direction which should reduce $\Phi$. Net magnetic induction shifts the $I_C (H)$
pattern away from zero field and reduces the zero field $I_C$
\cite{ryazanovquant}. Devices demagnetised both above and below
$T_C$ showed an increased value $R(0)$, above $R_{\mathrm{max}}$ as
expected \cite{pratt}, but no increase in zero field $I_C$ above $I_C^{\mathrm{max}}$. Hence it
would seem that stray induction does not strongly influence these devices.

In conclusion we have measured the Josephson current through a
pseudo-spin-valve structure. The zero field $I_C$ shows strong
modulation as the distinct remenant states of the PSV were mapped out. There was a strong
similarity between the $R(H)$ and zero field $I_C$. At
present theoretical work has been done on similar structures
\cite{bergeretprb, ChtchelJETP}, but in different limits to the
present work. A combination of these works may be required to explain the complex
behaviour due to the inhomogeneous magnetic structure in the PSV. 
It is nonetheless clear that $I_C$ can be actively controlled using the PSV
barrier.
If the barriers are replaced by weaker F layers, (e.g. 
Cu$_x$Ni$_{1-x}$ or Pd$_x$Ni$_{1-x}$) then for $T< 4.2$ K a much larger $J_C$ may
be large enough to produce spin torque effects \cite{WaintalPRB}. A suitable
design with the weak alloys and thicknesses chosen to be around the
$0-\pi$ transition should allow the creation a controllable
$\pi$-SQUID by switching one of the two devices into the 
anti-parallel state. This would
be a magnetic version of the voltage controlled
design using normal metal junctions \cite{baselmans2}. 
We acknowledge the support of the Engineering and Physical Sciences
Research Council, UK.

\end{document}